\documentclass[%
 reprint,
superscriptaddress,%groupedaddress,%unsortedaddress,%runinaddress,%frontmatterverbose,%preprint,%showpacs,preprintnumbers,%nofootinbib,%nobibnotes,%bibnotes,
amsmath,amssymb,
aps,
pra,
]{revtex4-1}
\usepackage{caption}
\usepackage{subcaption}
\usepackage{graphicx}
\usepackage{xcolor}
\usepackage{graphicx}% Include figure files
\usepackage{dcolumn}% Align table columns on decimal point
\usepackage{bm}% bold math
\usepackage{upgreek}
\usepackage{lineno}

\begin{document}

\preprint{APS/123-QED}

\title{Rotated chirped volume Bragg gratings for compact spectral analysis}

\author{Oussama Mhibik}
\affiliation{CREOL, The College of Optics \& Photonics, University of Central Florida, Orlando, FL 32816, USA}
\author{Murat Yessenov}
\thanks{yessenov@knights.ucf.edu}
\affiliation{CREOL, The College of Optics \& Photonics, University of Central Florida, Orlando, FL 32816, USA}
\author{Lam Mach}
\affiliation{CREOL, The College of Optics \& Photonics, University of Central Florida, Orlando, FL 32816, USA}
\author{Leonid Glebov}
\affiliation{CREOL, The College of Optics \& Photonics, University of Central Florida, Orlando, FL 32816, USA}
\author{Ayman F. Abouraddy}
\affiliation{CREOL, The College of Optics \& Photonics, University of Central Florida, Orlando, FL 32816, USA}
\author{Ivan Divliansky}
\thanks{corresponding author: ibd1@creol.ucf.edu}
\affiliation{CREOL, The College of Optics \& Photonics, University of Central Florida, Orlando, FL 32816, USA}

%\affil[1]{CREOL, The College of Optics \& Photonics, University of Central Florida, Orlando, FL 32816, USA}
%\affil[*]{Corresponding author: ibd1@creol.ucf.edu}

\begin{abstract}
We introduce a new optical component -- a rotated chirped volume Bragg grating (r-CBG) -- that spatially resolves the spectrum of a normally incident light beam in a compact footprint and without the need for subsequent free-space propagation or collimation. Unlike conventional chirped Bragg volume gratings in which both the length and width of the device must be increased to increase the bandwidth, by rotating the Bragg structure we sever the link between the length and width of an r-CBG, leading to a significantly reduced device footprint for the same bandwidth. We fabricate and characterize such a device in multiple spectral windows, we study its spectral resolution, and confirm that a pair of cascaded r-CBGs can resolve and then recombine the spectrum. Such a device can lead to ultra-compact spectrometers and pulse modulators.
\end{abstract}

%\setboolean{displaycopyright}{true}

%\begin{document}

\maketitle

Miniaturizing optical components has been a dominant research theme for the past few decades. However, such miniaturization is no guarantee for reduction of the overall size of an optical system performing a particular task \cite{Miller07JOSABFundamental}. This has been highlighted recently in the context of optical imaging, whereby replacing a conventional lens with an ultrathin metasurface \cite{Khorasaninejad16Science} does \textit{not} eliminate the free-space propagation distance required for image formation \cite{Guo20OpticaSqueezing,Reshef21NC}, thus leaving the volume of the imaging system undiminished. As a general principle, a minimum volume is required (for a given realizable refractive-index contrast) to achieve any task defined in terms of separating optical `modes' in a given basis \cite{Miller07JOSABFundamental,Miller19AOP}; in other words, optical devices ultimately need `thickness' \cite{Miller22Thickness}.  

\begin{figure}[t!]
    \centering
    \includegraphics[width=8.6cm]{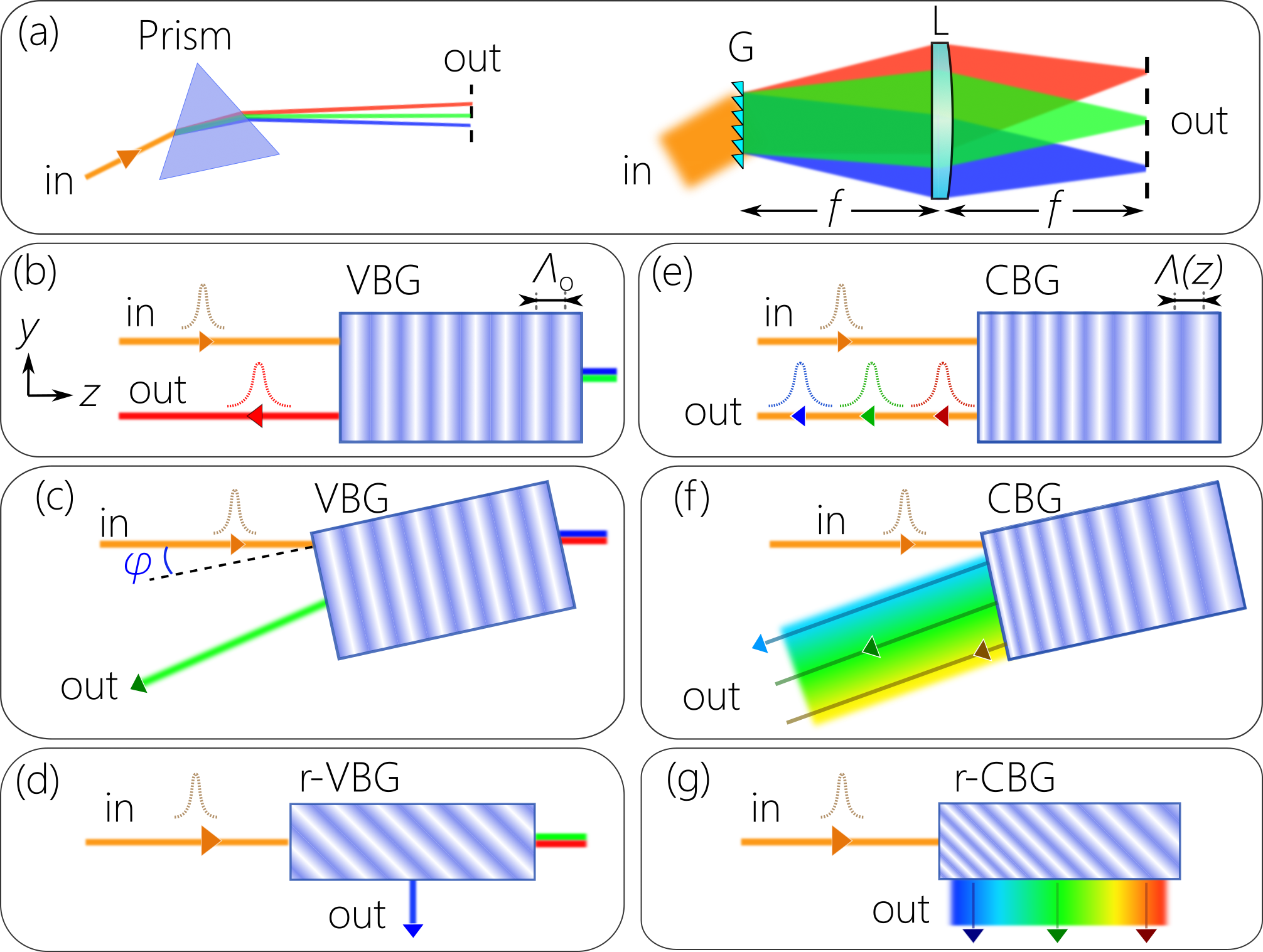}
    \caption{(a) Prisms and diffraction gratings spatially resolve the spectrum of incoming light. G: Diffraction grating; L: lens. (b) A resonant wavelength reflects when light is incident normally on a VBG, which (c) is shorter at oblique incidence. (d) When the VBG structure is rotated by $45^{\circ}$ with respect to the plane-parallel surfaces (r-VBG), the resonant wavelength exits from a facet orthogonal to the input. (e) A reflected pulse is stretched when incident normally on a CBG, and (f) its spectrum is spatially resolved at oblique incidence. (g) When the CBG structure is rotated by $45^{\circ}$ (r-CBG), the spectrally resolved wavefront emerges normally from the device.}%\vspace{-5mm}
    \label{fig:Concept}
\end{figure}

Another example is the ubiquitous task of spatially resolving the spectrum of light, which is at the heart of spectroscopy \cite{Hollas04Book}, ultrafast pulse modulation \cite{Weiner00RSI}, optical communications \cite{Gerken03IEEEPTL}, and in material identification, \cite{Wagatsuma21Book} for environmental \cite{Simonescu12}, biomedical \cite{Prasad03Book}, and chemical \cite{Henderson06Book} applications. Whether making use of a prism or a diffraction grating [Fig.~\ref{fig:Concept}(a)], or even a metasurface \cite{Arbabi17Optica}, the volume of the spectrometer is still dominated by free-space propagation, usually with the aid of a lens to spatially resolve and collimate the spectrum, which requires careful alignment and is sensitive to vibrations and shock.

The development of holographic recording in photosensitive glass has enabled the realization of photonic devices for spatially resolving the spectrum of an optical field; e.g., chirped Bragg volume gratings (CBGs) \cite{Glebov14OEng}. A pulse reflecting from a CBG at normal incidence is temporally stretched, and upon incidence on its opposite side, the stretched pulse is re-compressed \cite{Kaim14OEng}. Consequently, CBGs can be used in chirped pulse amplification systems, and have the advantage of a larger area with respect to fiber-based CBGs. Upon reflection at oblique incidence, on the other hand, the wavelengths reflecting from different depths within the CBG are laterally shifted with respect to each other (spatial dispersion \cite{Gerken03IEEEPTL}). Crucially, in contrast to a diffraction grating or prism, the spectrum is spatially resolved at the immediate exit from the CBG and thus does \textit{not} require additional free-space propagation. However, increasing the resolved bandwidth requires increasing the device length, which necessarily implies also increasing its transverse width to accommodate both the obliquely incident beam and the obliquely reflected spectrally resolved wavefront emerging from the same device facet. Consequently, the footprint of CBGs used for spectral analysis is quite large.

We propose here a new device that can alleviate this geometric constraint by severing the link between the length and width of the CBG-device used for spectral analysis, resulting in a significantly reduced device footprint. This is achieved by rotating the CBG structure by $45^{\circ}$ with respect to the plane-parallel facets of the device. We refer to such a device as a `rotated CBG' (r-CBG). In this configuration, the input field impinges normally on one facet of the r-CBG, and the spatially resolved spectrum also exits the device normally but from a \textit{different} facet that is orthogonal to the entrance facet. Crucially, the spectrum is resolved at the immediate exit of this facet, which can thus be readily integrated with other devices without the need for additional free-space propagation. We show that an r-CBG retains the spectrum-resolving performance of a CBG while providing the following advantages: reducing the device footprint; maintaining normal incidence onto and exit from the device; and separating the input and output beams in different paths. We produce here compact r-CBGs in the visible (centered at a wavelength of $\lambda_{\mathrm{o}}\!\sim\!580$~nm and bandwidth $\Delta\lambda\!\sim\!50$~nm), and near-infrared ($\lambda_{\mathrm{o}}\!\sim\!1$~$\upmu$m and $\Delta\lambda\!\sim\!34$~nm, and $\lambda_{\mathrm{o}}\!\sim\!0.8$~$\upmu$m and $\Delta\lambda\!\sim\!15$~nm). Such a device may impact spectroscopic applications by providing conveniently configured, easily aligned, small-footprint spectrometers and ultrafast pulse modulators.

We first consider broadband collimated light incident on a conventional volumetric Bragg grating (VBG) [Fig.~\ref{fig:Concept}(b)], which is recorded holographically in photosensitive glass by superposing two collimated beams at an appropriate relative angle. The reflected wavelength at normal incidence is $\lambda_{\mathrm{o}}\!=\!2n_{\mathrm{o}}\Lambda_{\mathrm{o}}$, and at oblique incidence the reflected wavelength is $\lambda(\varphi)\!=\!2\Lambda_{\mathrm{o}}\sqrt{n_{\mathrm{o}}^{2}-\sin^{2}\varphi}$, where $\Lambda_{\mathrm{o}}$ is the VBG period, $n_{\mathrm{o}}$ is its average refractive index (assuming a small index contrast), and $\varphi$ is the external incident angle [Fig.~\ref{fig:Concept}(c)]. Rotating the Bragg structure by $45^{\circ}$ with respect to the plane-parallel facets [Fig.~\ref{fig:Concept}(d)] results in a resonant wavelength of $\lambda_{\mathrm{o}}\!=\!\sqrt{2}n_{\mathrm{o}}\Lambda_{\mathrm{o}}$ exiting the device normally from a facet orthogonal to the input. To the best of our knowledge, such a `rotated VBG' (r-VBG) has not been reported to date. In general, VBGs are classified as transmissive \cite{Ciapurin06OEng} (when the resonant wavelength exits the facet opposing the input) or reflective \cite{Ciapurin12OEng} (when exiting from the input facet). The r-VBG configuration defies such a classification and occupies an intermediate position between these two types.

Consider next the conventional CBG illustrated in Fig.~\ref{fig:Concept}(e) with axially varying refractive index $n(z)\!=\!n_{\mathrm{o}}+\delta n\cos\{Qz+\beta(z-0.5L)^{2}\}$, where $n_{\mathrm{o}}$ is its average index, $\delta n$ its index contrast, $\beta$ its chirp rate, $L$ its axial length, and $Q\!\approx\!\tfrac{4\pi}{\lambda_{\mathrm{o}}}n_{\mathrm{o}}$. This grating structure is recorded holographically in photosensitive glass by superposing two beams, one diverging and the other converging. We ignore for simplicity chromatic dispersion in the grating material. Employing coupled-mode analysis \cite{Kogelnik69BSTJ,Kaim14OEng}, it can be shown that different wavelengths $\lambda$ are reflected from different depths $z$ within the CBG according to $\lambda(z)\!=\!\lambda_{\mathrm{o}}-\gamma z$, where $\lambda_{\mathrm{o}}\!=\!\lambda(0)$ and $\gamma\!=\!\tfrac{\lambda_{\mathrm{o}}^{2}}{2\pi n_{\mathrm{o}}}\beta$. Therefore, a normally incident collimated pulse is temporally stretched, which is useful in chirped pulse amplification systems.

\begin{figure}[t!]
    \centering
    \includegraphics[width=8.6cm]{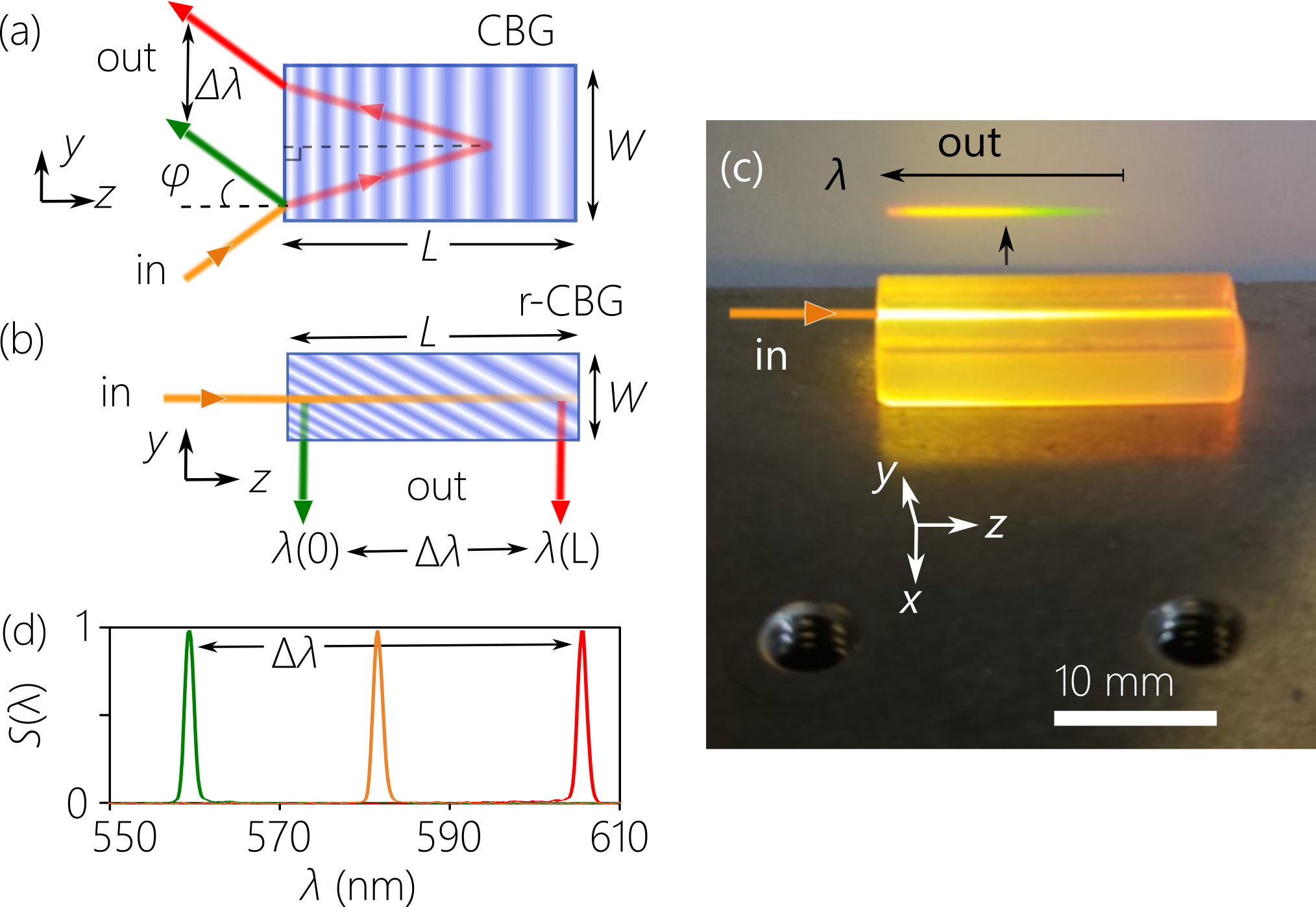}
    \caption{(a) Spatially resolving the spectrum of a beam obliquely incident on a CBG. (b) Proposed configuration for spatially resolving the spectrum of a beam normally incident on an r-CBG, and (c) a physical realization. (d) Measured output spectrum in (c) at $z\!=\!0, L/2,L$.}%\vspace{-3mm}
    \label{fig:Theory}
\end{figure}

At oblique incidence on the CBG, the spectrum of the reflected field is spatially resolved [Fig.~\ref{fig:Concept}(f) and Fig.~\ref{fig:Theory}(a)]. The resolved bandwidth is $\Delta\lambda\!=\!\tfrac{n_{\mathrm{o}}|\gamma|}{\sqrt{n_{\mathrm{o}}^{2}-\sin^{2}\varphi}}L$, where $\varphi$ is the external incident angle. The requisite length $L$ and width $W$ are:
\begin{equation}\label{eq:CBGWidthAndLength}
L=\frac{\Delta\lambda}{n_{\mathrm{o}}\gamma}\sqrt{n_{\mathrm{o}}^{2}-\sin^{2}\varphi},\;\;\;W=2\frac{\Delta\lambda}{n_{\mathrm{o}}|\gamma|}\sin\varphi,
\end{equation}
respectively. At normal incidence ($\varphi\!=\!0$), we have $W\!=\!0$, so that the spectrum is not resolved spatially. To increase the spatially resolved bandwidth $\Delta\lambda$ at oblique incidence, therefore, requires increasing \textit{both} $L$ and $W$. However, their ratio $\tfrac{W}{L}\!=\!2\tfrac{\sin\varphi}{\sqrt{n_{\mathrm{o}}^{2}-\sin^{2}\varphi}}$ is independent of $\Delta\lambda$ and depends on $\varphi$ alone. In other words, the two CBG dimensions $W$ and $L$ are \textit{not} independent, and instead go hand-in-hand: increasing the resolved bandwidth $\Delta\lambda$ requires increasing $L$, which in turn necessitates increasing $W$. Moreover, the CBG width $W$ in Eq.~\ref{eq:CBGWidthAndLength} is a lower limit, and in practice, it needs to be even larger due to two factors: (1) the incident beam width $\delta W$ increases $W$ by a factor $2\tfrac{\delta W}{\cos{\varphi}}$; and (2) a factor $2\tfrac{\sin\varphi}{\sqrt{n_{\mathrm{o}}^{2}-\sin^{2}\varphi}}\delta L$ must be added if the grating starts at a depth $\delta L$ below the surface.

In the case of an r-CBG [Fig.~\ref{fig:Concept}(g) and Fig.~\ref{fig:Theory}(b)], the connection between $W$ and $L$ is severed, so that we can continue to increase $L$ to increase the resolved bandwidth $\Delta\lambda$ \textit{while keeping $W$ fixed}. The only constraint on $W$ is the diffraction of the incident beam as it propagates along the r-CBG. The input beam is incident normally on an r-CBG facet, but effectively $\varphi\!=\!45^{\circ}$ because of the rotated profile index variation, so that $L\!=\!\tfrac{1}{|\gamma|}\sqrt{1-\tfrac{1}{2n_{\mathrm{o}}^{2}}}\Delta\lambda$, with $\Delta\lambda\!\approx\!1.13|\gamma|L$ at $n_{\mathrm{o}}\!=\!1.5$. Thus, by extending the r-CBG length $L$, we can resolve an ever-increasing bandwidth $\Delta\lambda$.

\begin{figure}[t!]
    \centering
    \includegraphics[width=8.6cm]{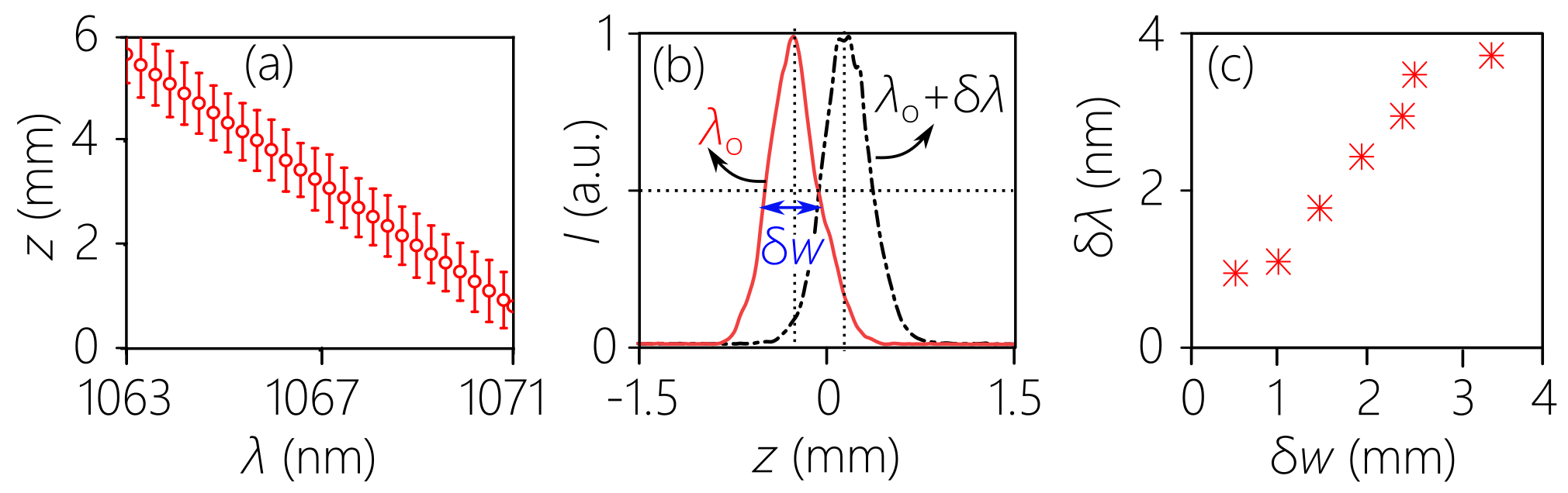}
    \caption{(a) Measured spatially resolved spectrum from an r-CBG. (b) Spatially resolving two wavelengths to identify the spectral resolution $\delta\lambda$ for an input beam size of $\delta W\!\approx\!1$~mm. (c) Dependence of the spectral resolution $\delta\lambda$ on beam size $\delta w$.}%\vspace{-5mm}
    \label{fig:Measurements}
\end{figure}

To visualize the performance of an r-CBG, we first fabricate one that operates in the visible spectrum, as depicted in Fig.~\ref{fig:Theory}(c), where the spectrally resolved beam ranges from green to orange. This r-CBG has dimensions 25$\times$12.5$\times$3~mm$^{3}$ ($L\times W\times H$) and was recorded using a set of convergent/divergent lenses of focal lengths 250~mm and -250~mm, respectively. The r-CBG operates at a central wavelength of $\lambda_{\mathrm{o}}\!\approx\!580$~nm and has a total bandwidth of $\Delta\lambda\!\approx\!49$~nm. To ascertain its performance as a spectral analyzer, we made use of a collimated white-light beam from a supercontinuum source (SuperK COMPACT, NKT Inc.). The beam width is $\approx\!1$~mm (FWHM), and was incident normally on the r-CBG input facet (dimensions $W\times H\!=\!12.5\times3$~mm$^{2}$), and the resolved spectrum exits normally from the orthogonal facet (dimensions $L\times H\!=\!25\times3$~mm$^{2}$), as shown in Fig.~\ref{fig:Theory}(c). To measure the spatially stretched spectrum, we collect light with a multimode fiber of core diameter 100~$\upmu$m connected to a spectrometer (Thorlabs, CCS175). We plot in Fig.~\ref{fig:Theory}(d) the spectra captured in this way at the beginning ($z\!=\!0$), mid-way ($z\!=\!L/2\!=\!12.5$~mm), and end ($z\!=\!L\!=\!25$~mm) points along $z$. The detected wavelength $\lambda$ varies linearly with $z$ as expected.

We carry out more detailed characterization in Fig.~\ref{fig:Measurements} on a second r-CBG designed for operation in the near infrared. This r-CBG has dimensions 25$\times$12.5$\times$6~mm$^{3}$ ($L\times W\times H$) and was recorded using a set of convergent/divergent lenses of focal lengths 1000~mm and -1000~mm, respectively. The r-CBG operates at a central wavelength of $\lambda_{\mathrm{o}}\!\approx\!1064$~nm and has a total bandwidth of $\Delta\lambda\!\approx\!34$~nm. A collimated beam of transverse width 2~mm (FWHM) from a tunable narrow-linewidth laser-diode source with a tunable wavelength range of $1055-1071$~nm, 10-pm tunning step, and 800-MHz linewidth (Velocity TLB 6121-H, New Focus) was normally incident on the r-CBG input facet. While tuning the input wavelength, the field exiting normally the r-CBG output facet is displaced axially along the $z$-axis. We capture the wavelength-tuned output beam displacement with a beam profiler (BladeCam-XHR, DataRay inc.) that registers the position and size of the output beam. We plot in Fig.~\ref{fig:Measurements}(a) this change in the axial position of the output beam over a bandwidth of $\approx\!8$~nm, which  varies linearly with the tuned wavelength, thereby confirming the expected linear chirp of the r-CBG.

We define the \textit{spectral resolution} of the r-CBG as the minimum wavelength separation $\delta\lambda$ that can be discerned at the output [Fig.~\ref{fig:Measurements}(b)], which increases with beam width $\delta W$ along the r-CBG $W$ dimension. We confirm this prediction by measuring $\delta\lambda$ while varying the beam width $\delta W$. For each beam width, the input wavelength was tuned until the intensity peak was displaced and the intensity dropped by $50\%$ at the original peak position [Fig.~\ref{fig:Measurements}(b)]. We varied $\delta W$ from 4~mm ($\delta\lambda\!\approx\!4$~nm) down to $\approx\!600$~$\upmu$m ($\delta\lambda\!\approx\!1$~nm), and observed the expected linear correlation between $\delta W$ and $\delta\lambda$. Further reduction in $\delta\lambda$ for high-resolution spectroscopic applications requires reducing $\delta W$ further. We comment below on this prospect.

\begin{figure}[t!]
    \centering
    \includegraphics[width=8.6cm]{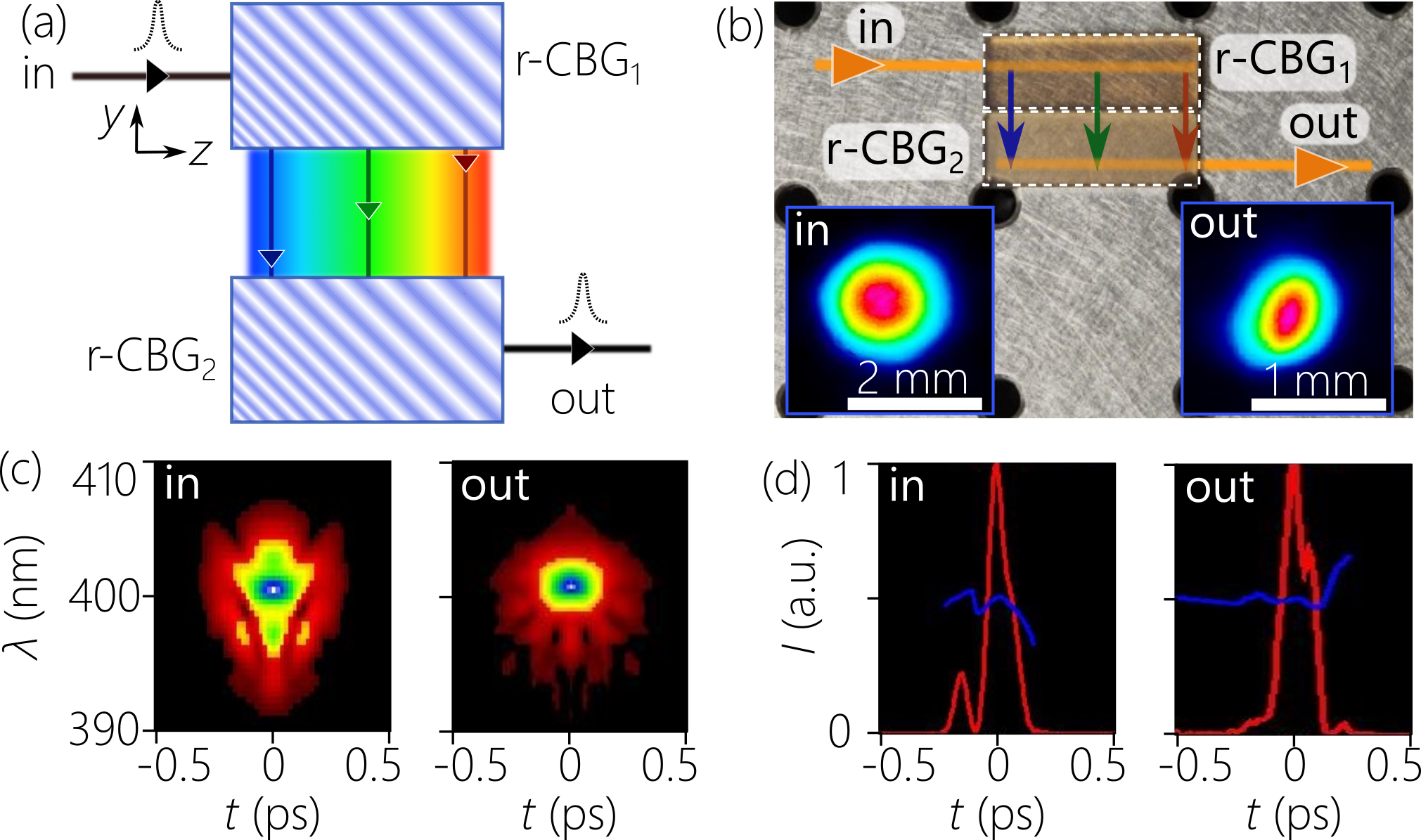}
    \caption{(a) Schematic of cascaded r-CBGs that resolve and then recombine the spectrum. (b) Photograph of a pair of cascaded r-CBGs as in (a), and the measured input and output beam profiles. (c) Measured FROG traces and (d) pulse profiles for the input and output in (b).}%\vspace{-5mm}
    \label{fig:rCBGpair}
\end{figure}

A cascade of two appropriately oriented identical r-CBGs will resolve and then combine the spectrum of the input field [Fig.~\ref{fig:rCBGpair}(a)]. To guarantee that the two r-CBGs are identical, we wrote a single r-CBG with dimensions 25$\times$12.5$\times$6~mm$^{3}$ ($L\times W\times H$) recorded using a pair of convergent/divergent lenses with focal lengths 1000~mm and $-1000$~mm, respectively. The r-CBG operates at $\lambda_{\mathrm{o}}\!\approx\!800$~nm and has a bandwidth $\delta\lambda\!\approx\!15$~nm. We then cut the r-CBG along the $H$ dimension to produce a pair with dimensions 25$\times$12.5$\times$3~mm$^{3}$. The cascaded r-CBG pair is depicted in Fig.~\ref{fig:rCBGpair}(b), where we confirm that the output beam size is similar to that of the input.

\begin{figure*}[t!]
    \centering
    \includegraphics[width=11.6cm]{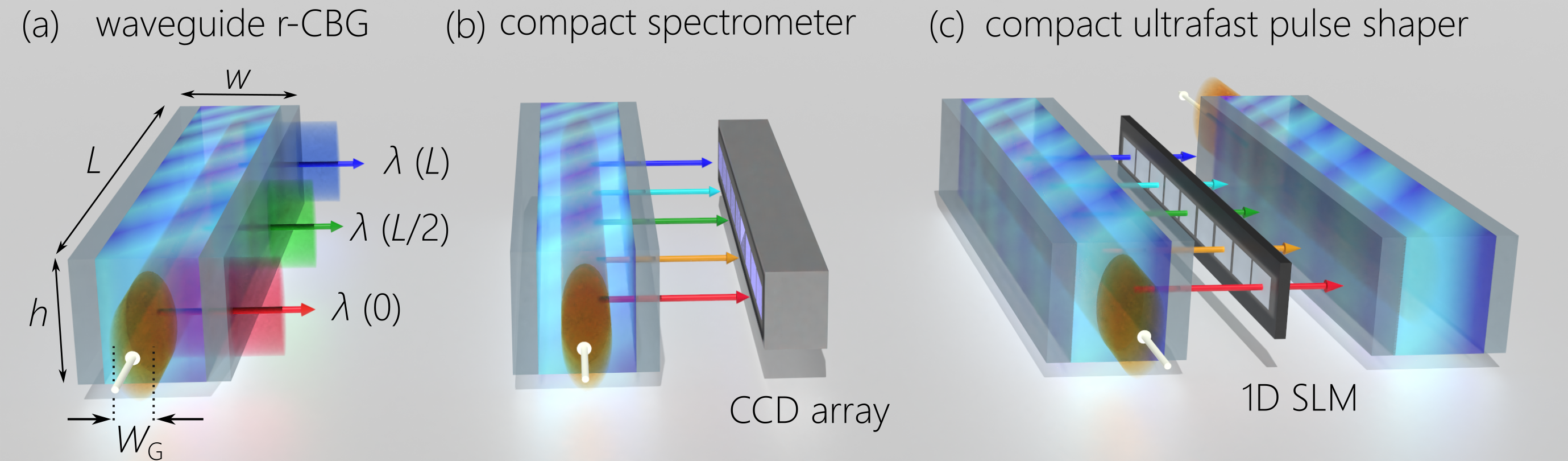}
    \caption{(a) Waveguide structure overlaid on the r-CBG to improve its spectral resolution. (b,c) Potential applications that benefit from the compact footprint of spectral analysis using r-CBGs: (b) compact spectrometer, and (c) ultrafast pulse modulator.}%\vspace{-3mm}
    \label{fig:Applications}
\end{figure*}

However, in the case of a pulsed input, an r-CBG introduces delays between the separated wavelengths (according to the traveled distance along $z$), which must be compensated for by the second r-CBG to guarantee that the pulse at the output has the same structure at the input's. We confirm this by characterizing the pulse at the input and output using a FROG autocorrelator (GRENOUILLE 8-50, Swamp Optics). The retrieved FROG traces of the measured pulses [Fig.~\ref{fig:rCBGpair}(c)] reveal the pulsewidths of the input and output pulses as $\Delta\tau_{\mathrm{in}}\!\approx\!100$~fs and $\Delta\tau_{\mathrm{out}}\!\approx\!180$~fs, respectively [Fig.~\ref{fig:rCBGpair}(d)]. A slight broadening of the pulse passing through the cascade of r-CBGs may be caused by uneven polishing of the r-CBG surfaces and misalignment in the system. 

Because a monochromatic beam emerges from the r-CBG with spatial width $\delta W$, the spectral resolution is $\delta\lambda\!\sim\!2|\gamma|\delta W$ (a higher-order correction follows from the finite diffraction strength of the r-CBG, which we ignore here). Unlike a diffraction grating where improving the spectral resolution (reducing $\delta\lambda$) requires increasing the beam size at the grating, the opposite is the case for an r-CBG: improving the spectral resolution requires reducing the input beam size. However, reducing $\delta W$ leads to diffraction as the input beam travels along the r-CBG length $L$, so that the spectral resolution deteriorates along $z$.

Further theoretical work is needed to ascertain the limits on the spectral analysis performed by an r-CBG, specifically with respect to the impact of the index contrast on the r-CBG efficiency and spectral resolution, and potential limits stemming from the presence of trace higher-order chirp. On the experimental side, the most important task is to improve the spectral resolution $\delta\lambda$ at fixed bandwidth $\Delta\lambda$, and thus increase the ratio $\eta\!=\!\tfrac{\Delta\lambda}{\delta\lambda}$. This ratio is governed by the fundamental limit set by diffraction of the field along the r-CBG. Therefore, one potential avenue for reducing $\delta\lambda$ and increasing $\eta$ is to introduce a guiding structure in the transverse dimension [Fig.~\ref{fig:Applications}(a)]. This can be achieved in the holographic writing process by adding a focused line using a cylindrical lens. For example, adding an index-guiding structure of width $W_{\mathrm{G}}\!\sim\!50$~$\upmu$m in an r-CBG of length $L\!=\!25$~mm yields $\eta\!\sim\!500$, corresponding to a spectral resolution $\delta\lambda\!\sim\!0.2$~nm over a bandwidth $\Delta\lambda\!\sim\!100$~nm.

The size of such a spectrum-analyzing device can be $25\times2\times0.2$~mm$^{3}$, with no need for any further free-space propagation. Potential applications of r-CBGs that harness the compactified spectral analysis are sketched in Fig.~\ref{fig:Applications}(b,c). First, an ultra-compact spectrometer can be envisioned by combining the waveguide-enhanced r-CBG with a 1D CCD chip [Fig.~\ref{fig:Applications}(b)]. The footprint of such a miniaturized spectrometer would be significantly smaller than diffraction-grating-based spectrometers. Second an ultrafast pulse modulator \cite{Weiner00RSI,Weiner09Book} can be envisioned as illustrated in Fig.~\ref{fig:Applications}(c), in which a 1D spatial light modulator (SLM) is sandwiched between two r-CBGs. By replacing the two diffraction gratings in the conventional $4f$ pulse modulator with two r-CBGs, the volume of the overall device is drastically reduced, its alignment is easier, and it can be more robust with respect to vibrations. Recently, combining this $4f$ pulse modulator arrangement with spatial modulation has led to the emergence of a host of novel spatio-temporally structured fields with unique features, including space-time wave packets (STWPs) \cite{Kondakci17NP,Kondakci19NC,Bhaduri19unpublished,Yessenov22NC}, transverse orbital angular momentum \cite{Hancock19Optica,Chong20NP,Gui21NP}, and toroidal pulses \cite{Wan22NP,Zdagkas22NP} (see \cite{Yessenov22AOP} for an overview). In all of these cases, relying on r-CBGs promises to compactify the experimental arrangements utilized.  We will present our results on the ultra-compact synthesis of STWPs elsewhere \cite{Yessenov23CompactSTWP}.

Our work here points to a more general strategy with respect to optical-device design. It is common to attempt miniaturization of individual optical components (e.g., a flat metasurface lens replacing a conventional finite-volume refractive lens). Of course, replacing an optical component with a flat surface reduces the volume occupied. However, this may not necessarily reduce the overall volume of an optical system, which is usually dominated by free-space propagation rather than individual devices (e.g., diffraction-grating-based spectrometers). Counter-intuitively, moving in the opposite direction by making use of \textit{volumetric} devices can instead reduce the system volume when the device combines two functionalities: 3D index variation and free propagation to separate the target modes. %We term this more general strategy `holomorphic optics'. (???)

In conclusion, we have fabricated and characterized a novel optical component, a rotated chirped Bragg volume grating (r-CBG), in which the multi-layer Bragg structure is rotated by $45^{\circ}$ with respect to the plane-parallel surfaces of the device. Consequently, only the length of the device -- and not its width -- needs to be increased for larger resolved bandwidths, in contrast to conventional CBGs in which both the length and width must be increased. The input beam is normally incident on the r-CBG, and the resolved spectrum emerges normally from an orthogonal facet, with no need for subsequent free-space propagation. Such a component can be a building block in compact spectrometers and ultrafast pulse modulators.

%\newpage

%\begin{backmatter}
\textbf{Funding}\\
U.S. Office of Naval Research (ONR) N00014-17-1-2458 and N00014-20-1-2789.\\

\textbf{Disclosures}\\
The authors declare no conflicts of interest.\\

%\bigskip
%\end{backmatter}

\bibliography{diffraction}
%\bibliographyfullrefs{diffraction}

\end{document}